\documentclass[aps,twocolumn,superscriptaddress,prl,floatfix]{revtex4}
\usepackage{amssymb}
\usepackage{amsmath}
\usepackage{graphicx}
\usepackage{xfrac}

\def\Idc{I_{\mathrm{DC}}}
\def\Ic{I_{\mathrm{C}}}
\def\icrn{I_{\mathrm{C}}R_{\mathrm{N}}}

\begin{document}
\title{Weak-link Josephson junctions made from topological crystalline insulators}
\author{R. A. Snyder}
\affiliation{Department of Physics, Joint Quantum Institute and the Center for Nanophysics and Advanced Materials, University of Maryland, College Park, MD 20742, USA}
\author{C. J. Trimble}
\affiliation{Department of Physics, Joint Quantum Institute and the Center for Nanophysics and Advanced Materials, University of Maryland, College Park, MD 20742, USA}
\author{C. C. Rong}
\affiliation{Army Research Laboratory, Adelphi, MD 20783, USA}
\author{P. A. Folkes}
\affiliation{Army Research Laboratory, Adelphi, MD 20783, USA}
\author{P. J. Taylor}
\affiliation{Army Research Laboratory, Adelphi, MD 20783, USA}
\author{J. R. Williams}
\affiliation{Department of Physics, Joint Quantum Institute and the Center for Nanophysics and Advanced Materials, University of Maryland, College Park, MD 20742, USA}

\date{\today}

\begin{abstract}
We report on the fabrication of Josephson junctions using the topological crystalline insulator Pb$_{0.5}$Sn$_{0.5}$Te as the weak link.  The properties of these junctions are characterized and compared to those fabricated with weak links of PbTe, a similar material yet topologically trivial. Most striking is the difference in the AC Josephson effect: junctions made with Pb$_{0.5}$Sn$_{0.5}$Te exhibit rich subharmonic structure consistent with a skewed current-phase relation. This structure is absent in junctions fabricated from PbTe. A discussion is given on the origin of this effect as an indication of novel behavior arising from the topologically nontrivial surface state.
\end{abstract}
\maketitle

Topological superconductors offer a new platform in which to study nontrivial ground states of matter. Since the early theoretical work of Read and Green~\cite{Read00}, and Kitaev~\cite{Kitaev01}, there has been a rapid expansion in the number of topological systems that possess superconducting correlations. Key in the investigation of topological superconductors is the tantalizing prospect of the creation and manipulation of Majorana fermions -- particles possessing non-abelian statistics that may prove useful in quantum computation. Experimental work has focused on topological superconductors created from proximitized one-dimensional nanowires with strong spin-orbit interactions~\cite{Mourik12} and time-reversal invariant topological insulators with either intrinsic~\cite{Sasaki11} or proximity-induced superconducting correlations~\cite{Williams12, Veldhorst12, Wiedenmann16, Deacon17}. Yet the list of possible topological superconductors does not end there and it is important to characterize these materials and elucidate the (potentially useful) differences therein.

\begin{figure}[b]
\center \label{fig1}
\includegraphics[width=3 in]{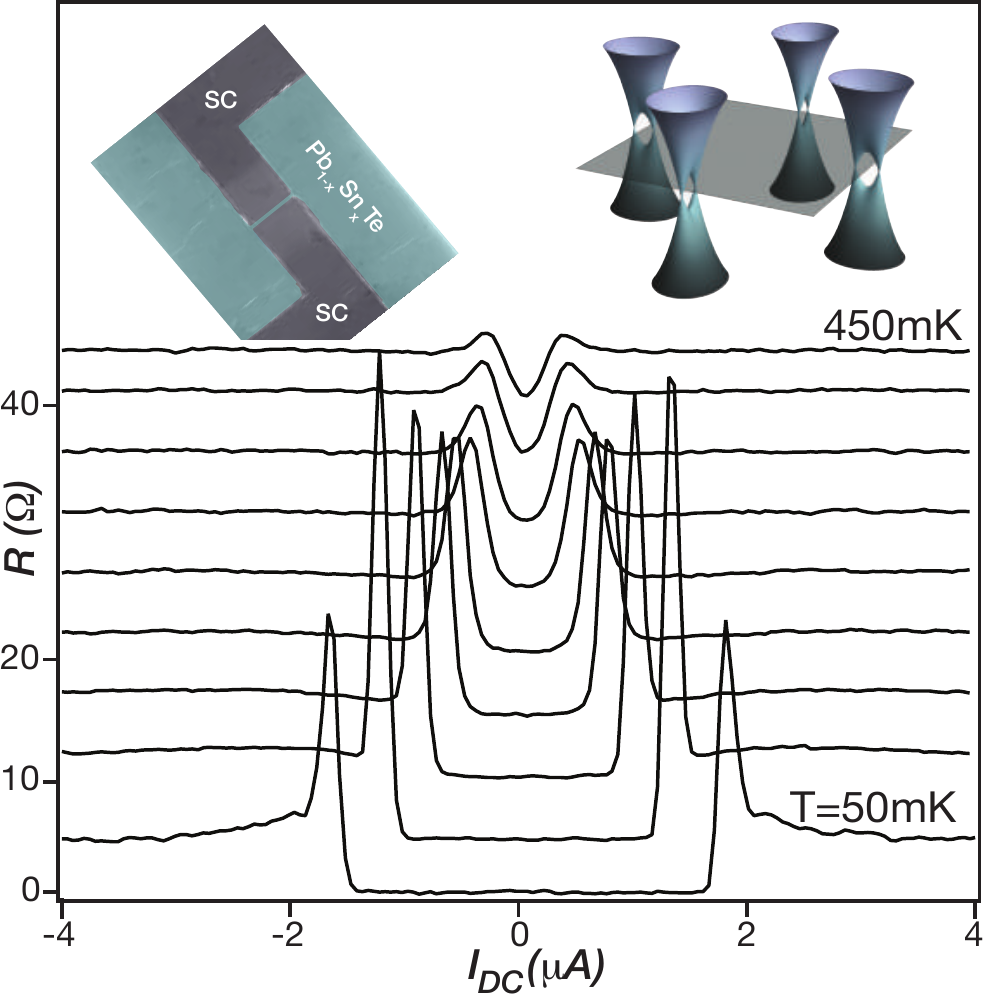}
\caption{\footnotesize{Temperature dependence of the differential resistance $R$, where superconducting features appear below $T$=500\,mK. The peaks in $R$ occur at values $\Idc=\Ic$. (Inset, upper left) Scanning electron micrograph of a device similar to the ones studied in this Letter showing two superconducting (SC) aluminum leads (dark grey) and the TCI material Pb$_{1-x}$Sn$_x$Te (green). Scale bar shown in white is 1$\mu$m. The spacing between the two SC leads is 100\,nm. (Inset, upper right) Schematic of the band structure of Pb$_{1-x}$Sn$_x$Te where 4 Dirac cones appear across the $\overline{X}$ point in k-space~\cite{Hsieh12}.}}
\end{figure}

Topological crystalline insulators (TCIs) produce topological states arising from the preservation of crystal symmetry~\cite{Fu11}. One of the first theoretically predicted TCIs was SnTe, and a band structure with 4 Dirac cones per unit cell was calculated (see the upper right inset of Fig. 1 for a diagram of the band structure) ~\cite{Hsieh12}. Soon after, experiments were able to demonstrate the topological nature of the surface state in SnTe and its cousin, Pb$_{1-x}$Sn$_x$Te, via angle-resolved photoelectron spectroscopy~\cite{Xu12, Tanaka13} and scanning tunneling spectroscopy~\cite{Okada13}. Important in these early experiments was the investigation of the topological phase transition in Pb$_{1-x}$Sn$_x$Se and Pb$_{1-x}$Sn$_x$Te as a function of $x$: a topological phase is observed for for values of $x>0.25$ in Pb$_{1-x}$Sn$_x$Se~\cite{Zeljkovic15} and for $x\ge0.4$ in Pb$_{1-x}$Sn$_x$Te~\cite{Xu12}.  More recently, theoretical investigations of the role of crystal symmetry in topological superconductors have begun. In particular, it was found that pairs of Majorana bound states can form and it was shown that a new class of topological superconductor in TCIs is possible~\cite{Liu14a, Fang14}. Experiments have shown superconductivity in In-doped SnTe, and odd-parity pairing indicative of a topological superconducting state has been observed~\cite{Sasaki12}. Superconductivity has been induced by the proximity effect, and SQUID circuits from topological crystalline superconductors have been fabricated with conventional SQUID behavior measured~\cite{Klett17} -- a reminder that transport from the trivial bulk may mask any novel behavior of the surface state. 

\begin{figure}[b]
\center \label{fig2}
\includegraphics[width=3 in]{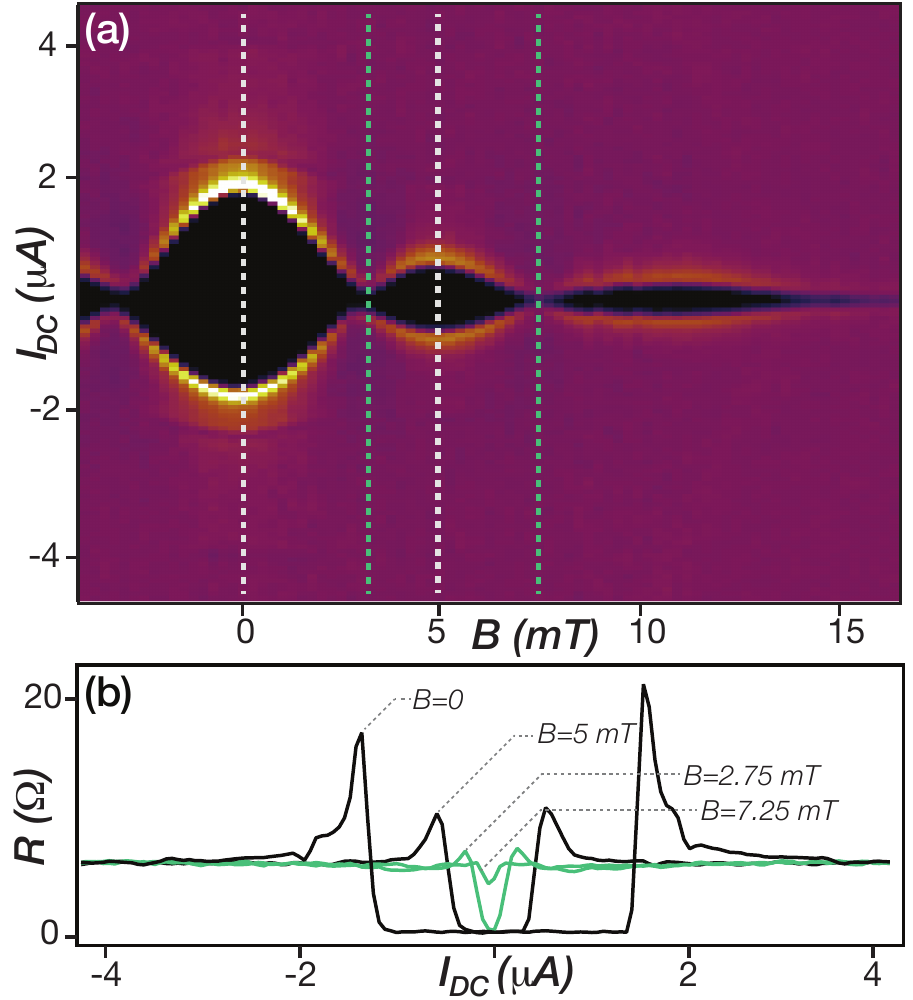}
\caption{\footnotesize{(a) Plot of $R(B,\Idc)$ revealing a Fraunhofer-like pattern consistent with a (nearly) uniform supercurrent across the width of the device. (b) One-dimensional cuts in the data from (a) at $B$=0, 5.00\,mT (black) and 2.75, 7.25\,mT (green), where the latter two show the variation in $R$ between at the first and second minimum in $\Ic$.}}
\end{figure}

Here we report on the fabrication of Josephson junctions using both Pb$_{1-x}$Sn$_x$Te (topologically nontrivial) and PbTe (topological trivial) as a weak link material between the two aluminum leads. Characterization of these junctions is carried out, including measurements of (near) DC $I-V$ curves, the $\icrn$ product and its temperature dependence, magnetic diffraction pattern, and the AC Josephson effect. The most striking deviation between the topologically-trivial and nontrivial junctions occurs under microwave radiation: in addition to the Shapiro steps observed at DC voltage values of $nhf/2e$, TCI Josephson junctions also exhibit steps at fractional values, indicating a strongly nonsinusoidal current-phase relation (CPR). The existence of higher harmonics in the CPR is confirmed through numerical simulations of the AC Josephson effect using a resistively-shunted junction model. The subharmonic structure reported here is distinct for weak-link materials with low mobility and we discuss the origin of this phenomena in terms of topological one-dimensional states measured in this material.

(111) Pb$_{1-x}$Sn$_x$Te epitaxial films were grown by molecular beam epitaxy (MBE) on semi-insulating GaAs substrates.  The (111) orientation was selected for two reasons: it offers rich topological states on the surface that are symmetric with respect to the (110) mirror planes, and it enables strain relaxation from dislocation glide along inclined (100) planes. The growth conditions were such that the surface adopts a simple ($1\times1$) reconstruction throughout growth as demonstrated by the RHEED pattern taken along the [110] azimuth~\cite{SuppInfo}.  The obtained compositions were either pure PbTe or nominally Pb$_{0.5}$Sn$_{0.5}$Te so as to facilitate discrimination between trivial insulator effects and TCI effects that emerge beyond the band inversion that is well known to occur somewhere in the range of  $0.2<x<0.4$~\cite{Xu12}. In contrast to conventional MBE method that employs (PbTe, SnTe) compound sources~\cite{Holloway81, Harman02}, the composition of the materials of the present work was instead controlled using individual elemental sources (e.g., Pb, Sn and Te2) having $>$99.9999$\%$ purity, and the surfaces were remarkably specular~\cite{SuppInfo}. Basic electrical characterization of the material was performed where it was found that Pb$_{0.5}$Sn$_{0.5}$Te was heavily p-doped (2.28x10$^{19}$ cm$^{-3}$) and PbTe heavily n-doped (-8.27x10$^{19}$ cm$^{-3}$) and each with a relatively low mobility of 113 and 160 cm$^2$V$^{-1}$s$^{-1}$ respectively.  In depth information on the characterization of the material can be found in Ref.~\cite{SuppInfo, Taylor17}.  

Josephson junctions with a width of 1\,$\mu$m and length between 50 and 120\,nm are patterned using electron-beam lithography.   The deposition of the contacts forming the junction begins with an in-situ argon plasma etch for 60\,secs at 50\,W followed by sputtering of Ti/Al (3\,nm/ 70\,nm).  During the deposition of the aluminum, the substrate is heated to 100$^o$C, and it was found that the $\icrn$ product of the junctions could be tuned from zero for room temperature deposition of the contacts to the value observed below~\cite{SuppInfo}.  Pb$_{0.5}$Sn$_{0.5}$Te and PbTe are removed through a reactive ion etch of Ar/H2 (20:2) everywhere except underneath the Al, in between the Al (the Josephson junction), and in a 2$\mu$m region on the left and right side of the Al. An SEM image of a completed device is shown in the inset of Fig. 1. Devices are then cooled to temperatures down to 50mK and differential resistance $R=dV/dI$ is measured in a current-bias  configuration (I$_{bias}$ between 1-10nA) with a lock-in amplifier. A total of 14 junctions showing superconducting properties were measured, two of which were investigated with the detail demonstrated in this Letter, each producing similar results~\cite{SuppInfo}. Spectroscopy of the device is obtained by applying a DC current source ($\Idc$), and plots of $R$ vs. $\Idc$ at different temperatures $T$ are shown in Fig. 1. Peaks in $R(\Idc)$ determine the critical current of the junction. The $\icrn$ product (R$_N$ is the normal state resistance of the junction) rises from zero at $T$=500mK to $\sim$ 10$\mu$V at base temperature (Fig. 1). 

\begin{figure}[t!]
\center \label{fig3}
\includegraphics[width=3 in]{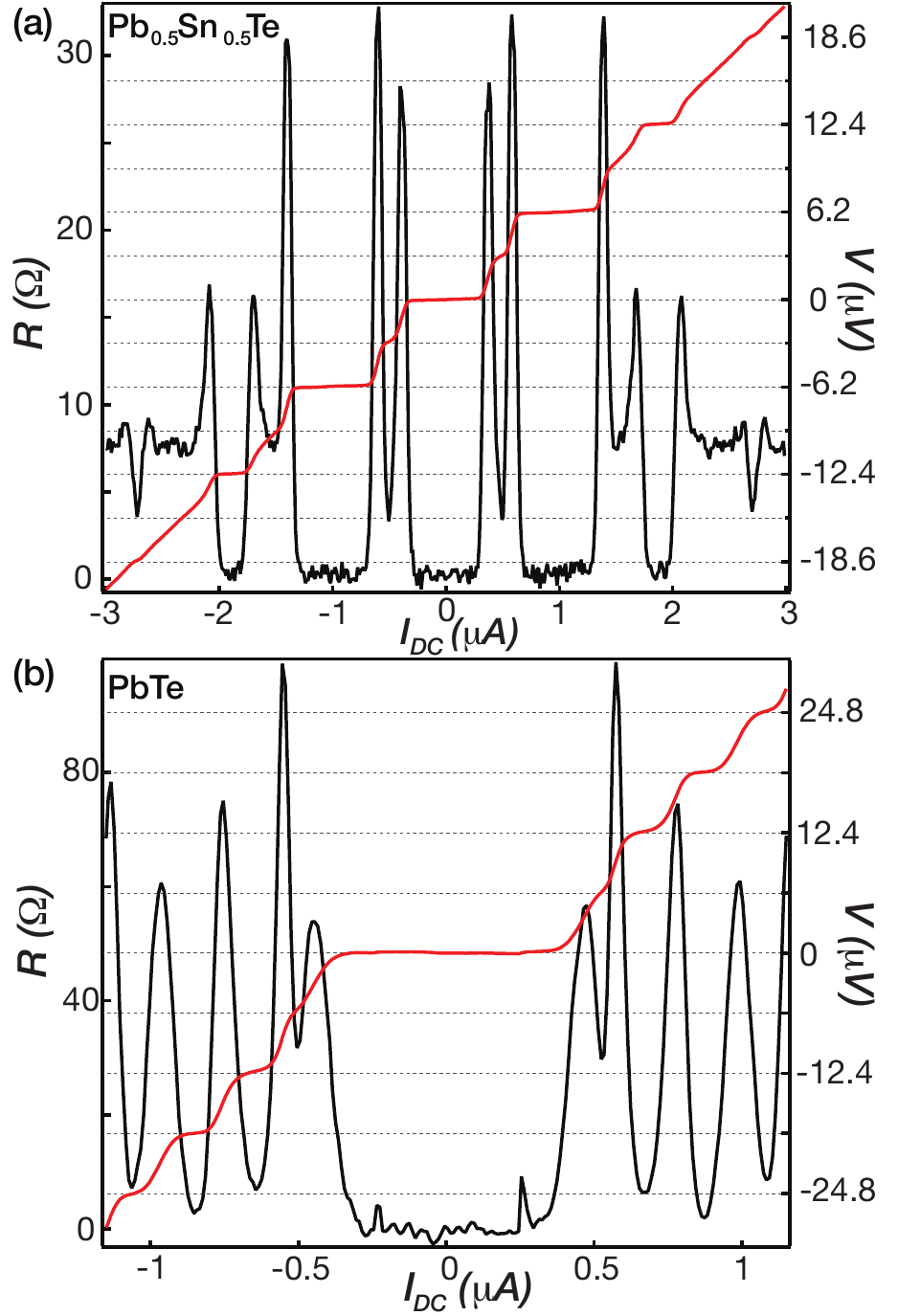}
\caption{\footnotesize{A comparison of Pb$_{0.5}$Sn$_{0.5}$Te and PbTe at $f$=3\,GHz and applied RF power of -6.75\,dBm. (a) Plot of $R$ showing minima at both expected values for Shapiro steps and at half integer values. Numerical integrated $I-V$ data (red) shows Shapiro steps at $nhf/2e$=n$\ast$6.2$\mu$V and additional features at fractional values of \sfrac{1}{2} and \sfrac{3}{2}. (b) By comparison, a PnTe device showing only integer values of the Shapiro steps, both in $R$ (black) and $I-V$ (red).}}
\end{figure}

Application of a perpendicular magnetic field $B$ allows for a variation of the superconducting phase difference along the width of the junction. A plot of $R$ as a function of $\Idc$ and perpendicular magnetic field is shown in Fig. 2(a). In conventional junctions with a sinusoidal CPR and a uniform magnitude of the supercurrent across the device, a Fraunhofer pattern in the magnetic-field dependence of $R$ is expected~\cite{Barone}. Importantly, Fraunhofer patterns or those resembling Fraunhofer patterns are experimentally useful for eliminating the possibility that the measured supercurrent in Fig. 1 arises from an electrical short between the superconducting leads, i.e. that the supercurrent is (at least approximately) uniform over the width of the device. Fig. 2(a) has a pattern that is reminiscent of a Fraunhofer pattern with two important deviations: the width of the central lobe is not twice the width of the other two, and while $\Ic \rightarrow 0$ at the second minimum (B=7.25\,mT), it remains finite at the first minimum (B=2.75\,mT). Cuts at B=2.75,5.00, and 7.25\,mT are compared with the $B=0$ plot in Fig. 2(b). This deviation of the magnetic-field dependence from a Fraunhofer pattern is consistent with other 3D topological insulators~\cite{Williams12, Veldhorst12}, and has been used in the past to imply nonsinusoidal current-phase relations~\cite{Kurter14}. However, a simple modification to allow for the critical current density to smoothly vary along the width of the device can also produce a similar modification of the Fraunhofer pattern. Hence, extraction of the current-phase relationship from measurements of this type can be tricky. Also visible in this plot is a small amount ($\sim$10$\%$) of hysteresis as a function of $\Idc$. Since the Stewart-McCumber parameter is small in junctions of this geometry~\cite{Oostinga13}, we ascribe this hysteresis to self heating of the electrons~\cite{LikharevRMP}.

\begin{figure*}[t!]
\center \label{fig4}
\includegraphics[width=6.75 in]{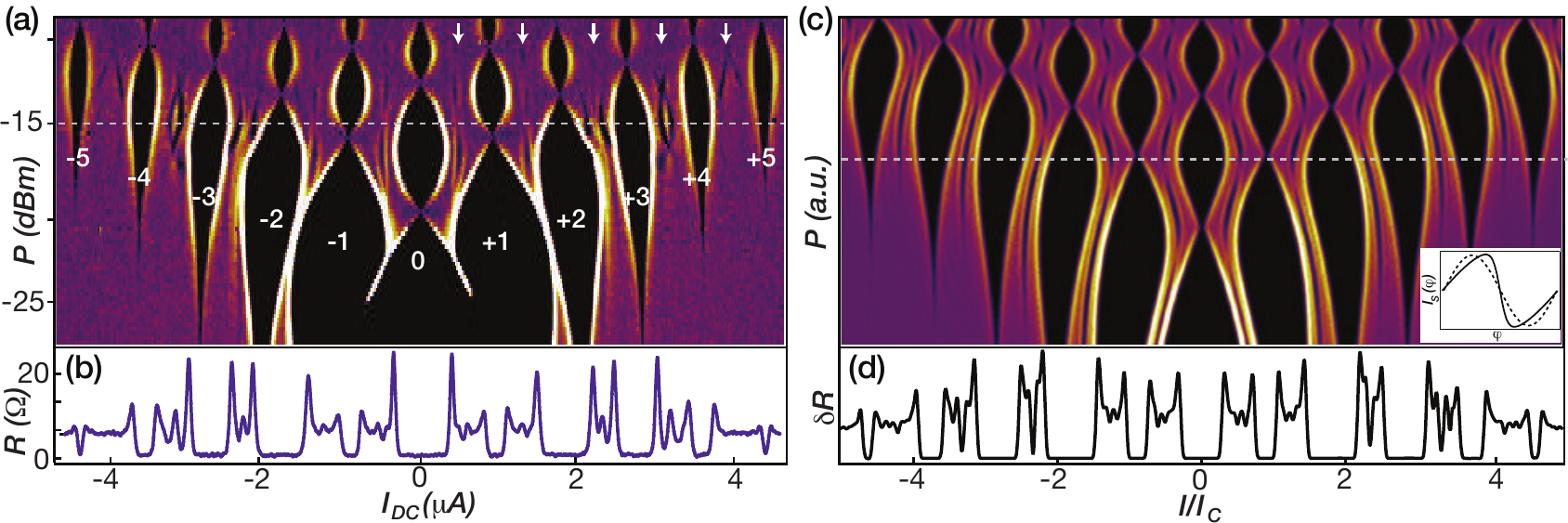}
\caption{\footnotesize{(a) Power dependence of the AC Josephson effect taken at $f$=2.2\,GHz and applied power range -27.25 to -9 dBm. In addition to the main Shapiro steps observed (black regions indicated by white numbers), structure in between the primary steps is measured (along the vertical indicated by white arrows). (b) A cut of $R$ taken along the grey line in Fig. 4(a). (c) Simulation of the RSJ model using a CPR for KO-2 theory for a ballistic Josephson junction, simulated over a similar range of parameters as the experimental data [Fig. 4(a)]. The saw-tooth behavior of the CPR is necessary to contain the higher harmonics in the CPS needed to mimic the experimental data. (Inset) A plot of the CPR for the KO-2 theory (solid line) is compared to a pure sine wave (dashed line). (d) A cut of the simulated differential resistance $\delta R$ qualitatively reproduces that observed in the experiment.}}
\end{figure*}

For a sinusoidal current-phase relation, a microwave voltage at frequency $f$ applied to the junction produces steps in the $I-V$ curves at voltages $nhf/2e$~\cite{Likharev}. These steps will appear as minima in the differential resistances $R$. Fig. 3(a) (black curve) shows $R$ vs $\Idc$ for an applied microwave frequency of 3\,GHz. Well-defined minima of $R$ are observed at values of $nhf/2e$. $I-V$ curves are generated from a numeric integration of the differential resistance (See Supp Info to view measured DC $I-V$ data, which also shows fractional steps). The steps in $I-V$ associated with these minima in $R$ are clearly seen in the generated $I-V$ data [Fig 3(a) red], corresponding to steps of $hf/2e$=6.2$\mu$V and in agreement with expectations. Besides these pronounced minima there is additional structure. Structure between the conventional minima is associated with higher harmonics of the CPR and enables the presence of fractional values of the AC Josephson effect.  The integrated $I-V$ shows a subharmonic feature at $hf/4e$ and $3hf/2e$, demonstrating that a modification of the conventional sin($\varphi$) CPR is observed in these junctions.

To investigate whether these half plateau steps arise from any topological properties of the weak link material  junctions were fabricated from the topologically-trivial material PbTe. $R$ under 3\,GHz radiation is presented in Fig. 3(b) (black curve), showing a conventional Shapiro step behavior with no structure in between plateaus. Also shown is the $I-V$ curve (red) showing only plateaus at multiples of 6.2$\mu$V, indicating that a current-phase relation arises primarily from a single frequency. Measurements of this junction at higher powers and frequencies show similar behavior: as a function of each, only integer Shapiro steps are observed~\cite{SuppInfo}. 

Further information on the CPR is revealed by a plot of the power dependence of the subharmonic structure. Fig. 4(a) is a plot of $R$ for an applied RF power $P$ between -27.25 and -9\,dBm, taken at $f$=2.2\,GHz. Fundamental frequency Shapiro steps are seen (labeled by number in white) and follow a Bessel function power dependence, as expected~\cite{Likharev}.  In addition, subharmonic structure is observed in between the primary plateaus (along the vertical indicated by the white arrows), with different structure between different Shapiro steps. For example, at certain values of $P$ a single dip is observed between steps 0 and 1, where two dips are seen between steps 3 an 4. This structure follows a more complicated pattern: as a function of power and $\Idc$ one and sometimes two minima are seen. A one-dimensional cut of $R$ [Fig. 4(b)] taken at P=-15dBm (grey line) demonstrates the intricate behavior observed in $R$. If only the fundamental and a second harmonic existed [$I_S \propto$ sin($\varphi$)+sin(2*$\varphi$)], only a strong, single dip in $R$ would be present between conventional Shapiro steps~\cite{SuppInfo}. This is not the case. 

\emph{Discussion}. As the AC Josephson effect data suggests, multiple harmonics are present in the CPR. Deviations from a sinusoidal CPR in low-capacitance weak link junctions are expected when the weak link has channels of high electron transmission~\cite{Golubov04}; recently this has been seen in one-dimensional nanowires with strong spin-orbit coupling~\cite{Spanton17}, graphene~\cite{English16}, and the three-dimensional topological insulator HgTe~\cite{Soch15}. Common to these three are the high values of the electronic mobility, with each experimental report citing highly-transmitting electronic channels as a source of the skewed CPR. This feature serves in stark contrast to the measured Hall mobility in the devices under study in this paper; for example, our reported mobility is $~\sim$250 times less than the reported mobility of the 3D TI HgTe~\cite{Soch15}.

To confirm that a current-phase relation possessing higher harmonics can mimic the data of Fig. 4(a), we perform a numerical integration of the resistively shunted junction (RSJ) model~\cite{Likharev} (see Ref.~\cite{SuppInfo} for details of the simulations). Theoretical predictions for the CPR as a function of transparency of the weak link have been made, where higher weak link transparency results in a more skewed CPR~\cite{SuppInfo, Golubov04, LikharevRMP}.  Fig. 4(c) is a simulation of the RSJ model using a current-phase relation with unity transparency:

\begin{equation}
I_S(\varphi)=\frac{\pi \Delta}{eR_N}\mathrm{sin}(\varphi/2)\mathrm{tanh}\frac{\Delta \mathrm{cos}(\varphi/2)}{2k_BT}.
\end{equation}

\noindent This CPR is shown in the inset of Fig. 4(c), plotted using an estimated value for $\Delta$ of $k_B\ast$500mK -- the temperature which the $\icrn$ product deviates from zero. While not all the features are captured by the simulation, a side-by-side comparison of the one-dimensional cuts of the experimental data and the simulation [Fig. 4(d), from cut taken along the grey line of Fig. 4(c)] shows a qualitative agreement, reproducing the essential features of the subharmonic structure. The most important distinguishing features of this CPR are the appearance of peaks centered between successive integer Shapiro steps and the unequal values of consecutive dips seen in the one-dimensional cut of the simulation [Fig. 4(d)]. These features are only observed in simulations with a strongly-skewed current-phase relation resulting from the existence of higher harmonics in $I_S(\varphi$)~\cite{SuppInfo}. However, a comprehensive search through various current-phase relations (like a CPR for diffusive systems)~\cite{SuppInfo} cannot account for all the experimental data. A feature that conventional CPRs fail to capture is fine structure in the power dependence of the subharmonic structure [see for example the region highlighted by the red box in Fig. 4(a)]. Whereas our data shows subharmonic lines crossing, simulations with conventional CPRs always have these lines running locally parallel. 

Recent experimental work investigating the surface of the TCI (Pb,Sn)Se has revealed one-dimensional, topological spin-filtered channels existing on step edges that break translational symmetry (called odd step edges)~\cite{Sessi16}. These 1D states only exist when the material is doped in the topological regime and theoretical arguments given in the manuscript indicate that this is a phenomena general to TCIs, not just the material under study. In fact, these 1D modes have been been observed in another TCI material, Bi2TeI~\cite{Avraham17}, and in a weak topological insulator Bi$_14$Rh$_3$I$_9$~\cite{Pauly15} -- of which TCIs are a subclass -- indicating that these modes are a general property in these types of materials. The existence of these states would account for the skewed CPR and the measured differences between Pb$_{1-x}$Sn$_x$Te and PbTe. The number of odd step edges in our samples can be estimated from the crystallographic offset of the GaAs wafer used to grow the Pb$_{1-x}$Sn$_x$Te, resulting in $\sim$5 per 100 nm. This number represents the minimum number, since steps edges can also be produced during growth. Each 1D mode is expected to contribute $\sfrac{e\Delta_0}{\hbar}$=10\,nA~\cite{Beenakker91}, so at least 500\,nA of the critical current could come from these modes. If only 500nA comes from the 1D modes, the rest will likely come from the bulk electrons, which will have a more conventional CPR. To check whether the subharmonic features survive this additional bulk supercurrent, we simulated a combination of a conventional sinusoidal CPR and the CPR from Eq. (1) in Ref.~\cite{SuppInfo} and the subharmonic features survive. Alternatively, the high electron transmission channel in the superconducting current of our PbSnTe junctions could be attributed to conduction in the two-dimensional (2D) topological surface states which is expected to have a much higher mobility than the measured bulk Hall mobility.  The expected Thouless energy for these 2D surface states is greater than $k_BT$.  This alternate explanation for the skewed CPR is somewhat similar to the analysis of similarly skewed CPR data from other work~\cite{Soch15}.

We thank Fred Wellstood for useful discussions and to C. Dickel, F. Luthi and the DiCarlo lab for assistance with measurements. This work was sponsored by the grants National Science Foundation Physics Frontier Center at the Joint Quantum Institute (PHY-1430094) and the Army Research Office (W911NF1710027). 

\nocite{S1}
\nocite{S2}
\nocite{S3}

\end{document}